# An Exploration Tool for Retrieval of Travel Information with Personal Photos

Risa Kitamura[1)]    Takayuki Itoh[1)]

1) Ochanomizu University

## ABSTRACT

Photos can be treated as life logs of photo owners. Photos can be reliable information to estimate patterns of actions and movements of the owners. Based on this discussion, we are developing an interactive technique to explore the recommended tourist spots based on their past personal travel photos. The technique extracts a set of keywords from the photo set applying a generic object recognition and constructs a tree structure to support the exploration of the keywords. When a user selects a set of interesting keywords, the system provides travel information related to the selected keywords. Our previous paper already introduced the visualizations that demonstrate the appropriateness of the structure of the keywords. This paper focuses on the mechanism for interactive travel information retrieval of our system and user evaluations with this system.

## 1  INTRODUCTION

Thanks to the recent evolution of digital cameras and mobile devices, we can freely take pictures and share them with our family or friends nowadays. We can analyze the preferences and behaviors of photo owners by collecting their photos as life logs. Based on this background, we are developing a travel information recommendation system based on the analysis of personal travel photos.

Figure 1 shows the processing flow of the system we are currently developing. This system provides an interactive information retrieval mechanism so that users can select query keywords related to their past travel. To realize this mechanism, the system extracts keywords from the users' past photos by applying a generic object recognition technique. Then, our technique generates a graph structure by treating the keywords as vertices and connecting pairs of vertices if they co-occur more than a user-defined number of photos. The technique converts the graph into a simplified tree structure where nodes of the tree correspond to clusters of keywords. The system provides a Web-based user interface that displays the structured keywords and related photos. We suppose that users select keywords by clicking nodes and photos on the Web browser and then specifies a region name. The system retrieves the travel information by treating the selected keywords as query words and specifying the region name, and finally displays the retrieved information on the Web browser. Here, we define "region" in this paper as a city or an area that travelers stay for one or several days, such as "Tokyo," Paris," "Alps," or "Hawaii."

One of the typical target use cases is that travelers who have taken many travel photos are going to retrieve travel information specifying a region they will visit. We expect the presented user interface enables users to remember their past travels and realizes enjoyable exploration of keywords that are appropriate as query words to retrieve information on their future travels.

There have been several studies on travel information recommendation applying past travel photos [2,3,4,8,9,11] as introduced in Section 2. However, most of these studies supposed to input GPS-enabled photos. We apply a genetic object recognition process to the input travel photos without GPS information and extract keywords related to past photos so that users can interactively select query keywords for travel information retrieval.

We have already presented the visualization of the graph structure of the keywords extracted from users' past photos [7] as introduced in Section 3. We present the user interface we developed after the previous study in Section 4, and discuss the usability of the user interface introducing our user experiments in Section 5.

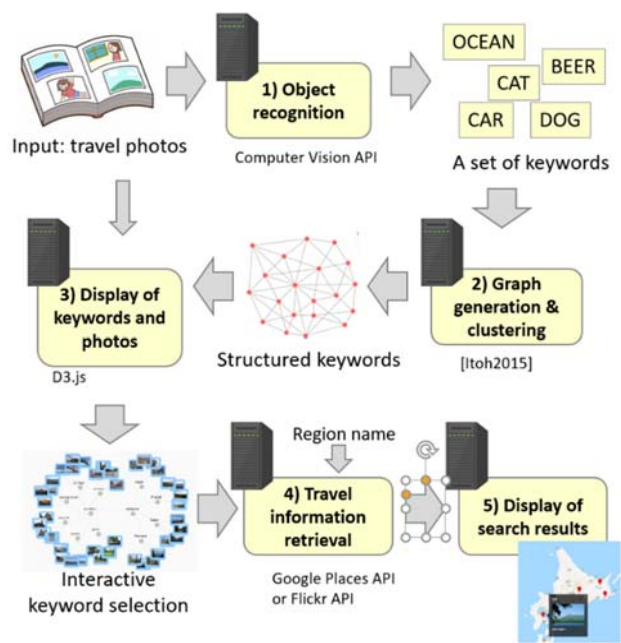

Figure 1: Processing flow of the presented system.

## 2  RELATED WORK

### 2.1  Interactive System for Retrieval of Travel Information

There have been several studies and developments on interactive travel information retrieval systems.

Kurashima et al. [8] presented a travel route recommendation technique applying GPS information of photos uploaded onto Flickr. The technique generates behavior models by tracing movements of persons with GPS information of photos. Cheng et al. [3] also generated behavior models while dividing photo owners on Flickr based on their attributes including age, gender, and race. W2Go presented by Gao et al. [4] suggests a ranked set of landmarks in a specific region by analyzing GPS information and annotations of photos on Flickr and travel information retrieved from Yahoo Travel Guide. The system presented by Cao et al. [2] retrieves clusters of photos taken at a specific region and their related keywords based on user-specified query keywords. Photo2Trip presented by Lu et al. [9] interactively recommends travel routes based on user-specified target regions, schedule, and favorite travel styles, by extracting famous landmarks from GPS-enabled photos on Panoramio and then retrieving information related to the landmarks from travel information Web sites.

These studies suppose GPS information is embedded in photos, and therefore, it is difficult to retrieve appropriate information if GPS information is not available in the input photos. On the contrary, our study does not require GPS information but applies a genetic object recognition process to extract keywords related to past travels of photo owners. Also, we developed an interactive mechanism to assist the selection of query keywords by displaying both photos and textual information while existing studies just provide a text input interface for the specification of query keywords.

## 2.2 Interactive Image Browser

The presented user interface displays multiple photos in a limited display space. The interactive image browser has been a well-studied research topic since it is helpful for interactive exploration of user-interested images from a large-scale image collection. Existing techniques on image browsers are divided into "structured" [1, 5] and "unstructured" [10, 12] techniques. The user interface presented in this paper is a kind of structured image browser since our technique constructs a tree structure of keywords and displays multiple photos associated with the keywords in the tree structure. The user interface displays the tree structure by "node-link" representation that displays the structures as lines connecting parent and children nodes, while many structured image browsers apply "space-filling" photo layout algorithms.

## 3 EXTRACTION AND STRUCTURING OF THE KEYWORD

This section introduces the implementation details on keyword extraction and graph generation [7].

### 3.1 Keyword retrieval

We use a generic object recognition function of Microsoft's Computer Vision API [13] to recognize subjects taken in past travel photos, and then assign the photos keywords corresponding to the subjects as tags. This function returns information on the visual content in the image such as objects, living beings, scenery, and actions. It also returns a confidence score that represents the confidence of a tag in real numbers between 0 and 1.

### 3.2 Graph construction and clustering

This paper denotes the confidence of the *j*-th keyword retrieved from the *i*-th photo as $c_{ij}$. Also, let the number of keywords *m*, and the number of photos *n*. Our implementation generates a graph G={V, E}, where V is a set of vertices, and E is a set of edges. The definition of a vertex and an edge in G is as follows:
- A vertex (corresponding to a keyword) has an *n*-dimensional vector including $c_{1j}$ to $c_{nj}$.
- An edge connects two vertices if the inner product between their vectors is larger than a user-specified threshold (0.1 in our implementation). It simply denotes that pairs of keywords that co-occur in the same photos are connected by edges.

Then, our implementation applies our own graph clustering algorithm [6] that puts a set of vertices into the same cluster if similarities of their vectors and adjacent nodes are larger. The algorithm brings clustering results that vertices connected to a larger number of adjacent vertices are separated from larger clusters. In other words, vertices separated from larger clusters correspond to high-level concept keywords that co-occur with various keywords in the travel photos. Also, some other vertices corresponding to low-level concept keywords that co-occur with the same high-level concept keyword belong to the same cluster. For example, this algorithm forms a cluster consisting of names of animals such as "cat" and "dog" and then connects this cluster to another cluster consisting of higher-level concept keywords such as "animal" and "mammalian."

Figures 2 and 3 demonstrate that the above algorithm appropriately constructs a graph of the keywords.

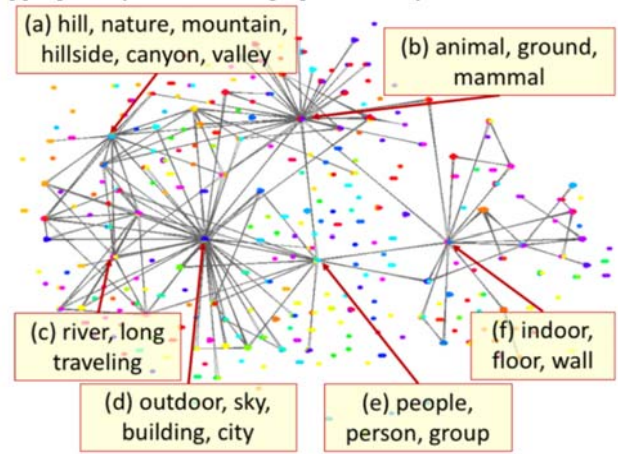

Figure 2: Hub nodes corresponding to commonly retrieved keywords.

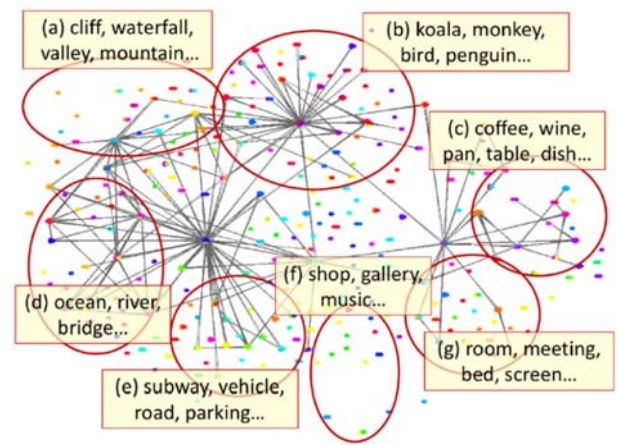

Figure 3: Typical keywords connected to the hob nodes.

## 4 USER INTERFACE FOR TRAVEL INFORMATION RETRIEVAL

This section presents the implementation details of our user interface for query keyword selection and display of retrieved travel information.

## 4.1 Tree Display for Query Keyword Selection

We developed a Web-based user interface to assist the query keyword selection for retrieval of travel information based on the users' past travels. It parses the graph structure introduced in the previous section described in JSON format, converts the input information into a simpler tree structure, and displays on the Web browser. Our user interface applies the graph visualization module "Force Layout" supplied by D3.js to display on the Web browser.

The paper defines the nodes of the simplified tree structure as follows:

Node: A cluster of vertices. Here, a node has one or more representative keywords selected based on their frequency in the photos.

Root node: one of the nodes which have the vertex connected by the largest number of edges.

Hub node: the nodes connected with the root node. Here, a node is treated as a hub node if it is related to the root node by more than a user-defined number of edges.

Figure 4(left) illustrates the definition of the tree structure.

The user interface displays the root node at the center of the display space and hub nodes around the root node. It also displays a representative keyword belonging to the root or a hub node that is most frequently appeared in the input photos. Furthermore, the system selects several photos related to the keywords contained in each of the nodes and displays the photos connecting by edges from the node. This paper calls the selected photos "photo node." Users can associate keywords of the node by looking at the selected photos.

The detailed processing flow of the photo node selection is as follows. First, our implementation calculates $score_i$ for the $i$-th keyword by the following equation:

$$score_i = conf_i / appear_i \quad (1)$$

Here, we suppose the confidence $conf_i$ and frequency $appear_i$ of the $i$-th keyword is calculated in advance. Then, the system calculates $\Sigma score_i$, the sum of the scores of keywords contained in the current node. Finally, it selects the photos in the order of the $\Sigma score_i$ and treats the selected photos as the photo nodes of the current node.

The user interface initially displays a simple tree structure consisting of the root and hub nodes, and photo nodes associated with the hub nodes. Other nodes are not displayed at this moment. When a user specifies a node by a click operation, the user interfaces additionally display the nodes connected to the clicked node. Figure 4(right) shows the design of the user interface.

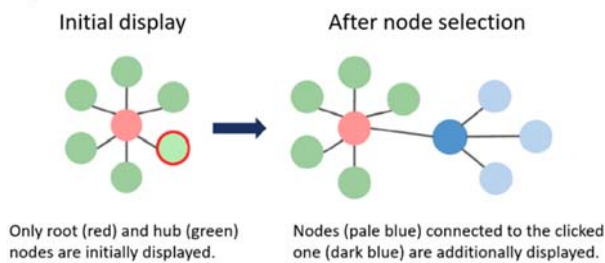

Figure 4: Definition and interaction with nodes of the tree structure.

The keywords associated with the nodes that have a larger number of edges are usually high-level concept keywords that co-occur with various keywords in the travel photos. Based on this principle, our user interface firstly displays only the root and hub nodes that contain high-level concept keywords. Users can freely select hub nodes corresponding to user-interested high-level concept keywords and display connected nodes corresponding to lower-level concept keywords. This mechanism is useful for the exploration and selection of keywords for the query of travel information. Meanwhile, users can eliminate hub-nodes that are out of their interest in the display space.

## 4.2 Query of Travel Information

Our user interface queries travel information with the region name and a set of keywords selected from the tree structure displayed on the Web browser.

Our implementation switches the following two APIs to retrieve the regional information.

Places API (Google Maps Platform)[14] is useful to retrieve information on region-related facilities and services. We applied "Text Search Requests" because it enables us to retrieve information by matching categories and keywords or ambiguous region names.

Flickr API[15] supports to retrieve photos by the matching between the metadata of photos and query words.

Our implementation retrieves the travel information by the following processes applying the "flickr.photos.search" method.

1. Retrieve photos from specific region names and query keywords.
2. Count the number of photos taken near the retrieved photos.
3. Treat the place that the retrieved photos are taken as a sightseeing place.
   - Relationship between query keywords and the metadata of photos is significantly high.
   - The number of photos taken near the retrieved photos is sufficiently large.
4. Retrieve the information related to the place.

Our user interface supposes that users interactively select nodes or photo nodes. The keywords associated with the selected nodes or photo nodes are used as query keywords to retrieve the information about the place.

Finally, our implementation displays the retrieved information onto Google Map by applying "Maps JavaScript API" [16].

## 5 RESULTS

We applied a travel photo set containing 2,581 photos taken during Japanese domestic and international travels of the male photo owner in this study. We applied Computer Vision API to extract keywords, constructed a graph structure of the extracted keywords, and displayed the simplified tree structure by our user interface on a Web browser. The graph structure constructed from this photo set is shown in our previous paper [7].

Figure 5 shows the initial display of the tree structure constructed by our technique. The Force Layout module implemented by D3.js places the root node at the center of the display space and hub nodes around the root node. The example displays four photos for each of the hub nodes. The system selects keywords associated with each of the hub nodes based on their frequency in the object recognition result and displayed around each of the hub nodes. When a user clicks a hub node, the system displays other keywords associated with the clicked hub node and frequently (five or more in our implementation) appeared in the object recognition result.

Figure 6 shows an example that a user clicked the hub node "nature" and then the system displays additional nodes connected to the clicked node. The example shows that lower-level concept keywords "forest" and "lake" related to the keyword "nature" are additionally displayed. The figure demonstrates that the system supports users to select higher-level concept keywords first and then suggests lower-level concept keywords so that users can narrow down their interests in their travel plans.

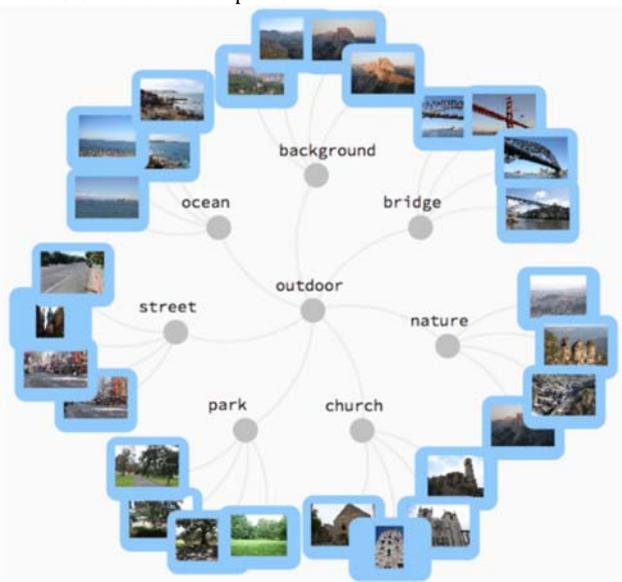

Figure 5: Initial display of the tree structure.

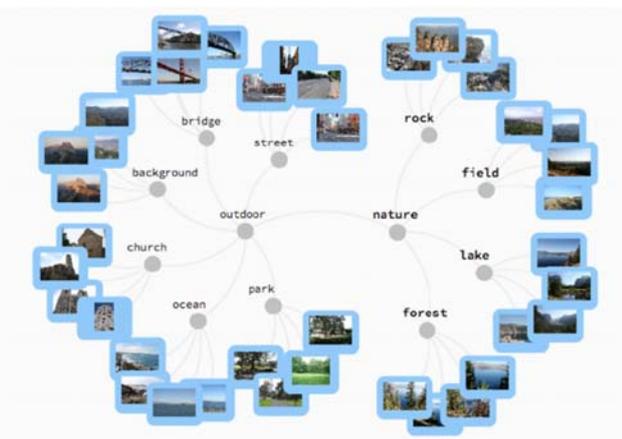

Figure 6: Click operation to additionally display lower-level concept keywords.

Figure 7 shows a snapshot of the user interface for the operation of the tree structure. The window displays a set of keywords related to a photo node specified by a mouse-hovering operation. Additional nodes connected to the hub node of the photo node are displayed as shown in Figure 6 when a user presses "SHOW CHILD NODE" button. Users can input arbitrary keywords when they press "SEARCH NODE BY KEYWORDS" button. Users can select nodes by pressing "SELECT NODE" button when they want to specify the query keywords. When a user presses "SEARCH SPOTS" button after specifying the query keywords while the above operations, the words are sent to the APIs for travel information retrieval. The window of the user interface is switched within one second for each of the above operations.

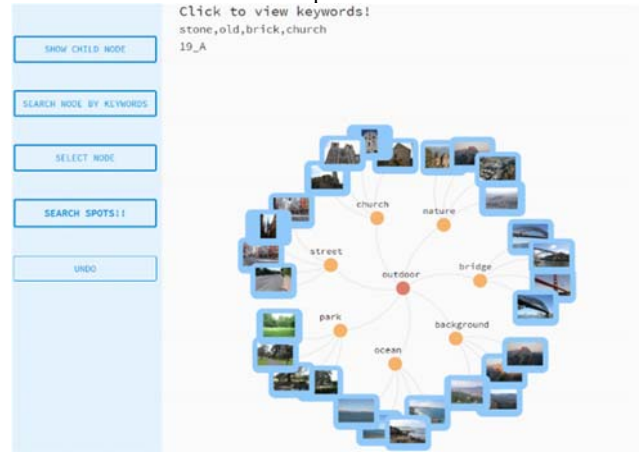

Figure 7: Our user interface running on the Web browser.

Figure 8 shows an example of the retrieved travel information. This example displays the places related to the retrieved information as markers on Google Map. Detailed information is displayed in the text area when a user specifies a marker by a click operation. This example is archived when a user specifies the region as "Hokkaido" and the keyword "lake" as shown in Figure 6.

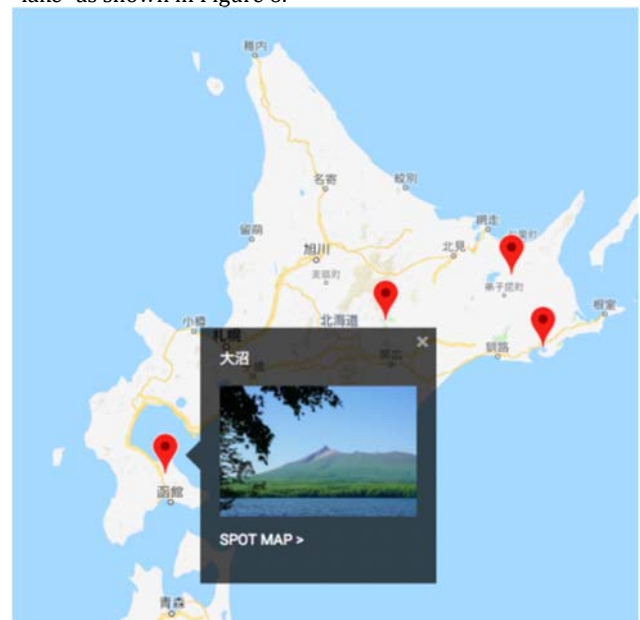

Figure 8: Example of retrieved travel information displayed on Google Map.

# 6 USER EVALUATION

We conducted user experiments applying the photo set introduced in the previous section. We invited ten participants who are female university students majoring in computer science to the first experiment. The owner of the photo set is invited to the second experiment.

## 6.1 Evaluation with participants

We asked participants to evaluate how they can easily find interesting information related to specific cities or regions by using the presented system.

Firstly, we asked to evaluate the appropriateness of the photo selection by applying the equation (1) with the keywords extracted by Computer Vision API. We prepared the following sets of photos:

<u>Photo-set A:</u> selected based on the scores calculated by equation (1).
<u>Photo-set B:</u> selected randomly.

We asked participants to play with the presented user interface and count the number of photos that they feel appropriately selected. The total number of displayed photos was 122. Table 1 shows the average numbers of the appropriate photos. The result shows that the photo selection by the presented technique was better than the baseline implementation.

Table 1. Evaluation of photo selection.

|  | Photo-set A | Photo-set B |
|---|---|---|
| Average appropriate photos | 88.0 | 68.7 |

Then, we had a user evaluation on recommendation results. Specifically, we experimented to observe the effectiveness of information retrieval based on keyword extracted by object recognition, satisfaction of the search results by the applied APIs, and effectiveness of switching the APIs. We applied two regions "Okinawa" and "Hokkaido" in this experiment, and asked participants to select one keyword and one photo and evaluate the recommended information in the five-point Likert scale. Here, we applied the following three APIs:

<u>API-set 1:</u> Places API (with the order of the score of user reviews)
<u>API-set 2:</u> Flickr API (with the order of the correlation with keywords)
<u>API-set 3:</u> Flickr API (with the order of the number of photos)

Table 2 shows the average of the evaluations. The result shows that Flicker API was more preferable our study. Also, it is possible that the order of suggested information based on the score of user reviews is not always effective in our study. Meanwhile, we could not find significant differences between API-sets 2 and 3. It suggests the meaningfulness of preparing multiple APIs.

Finally, we asked participants to evaluate the operability of the user interface and representation of the information recommendation results in the five-point Likert scale. Table 3 shows the average of the evaluations. This result shows that participants preferred our implementation. Most of the comments by the participants were related to desirable additional functions such as display of date and place of the photos, a suggestion of Web sites related to the sightseeing spots, and emphasis on user-selected photos. No participants complained about the usability and performance of the user interface in their comments.

Table 2. Average of the evaluations for three API-sets.

| API-set | 1 | 2 | 3 |
|---|---|---|---|
| Okinawa, Keyword | 2.0 | 4.2 | 4.5 |
| Okinawa, Photo | 2.8 | 3.9 | 3.7 |
| Hokkaido, Keyword | 1.7 | 4.1 | 3.9 |
| Hokkaido, Photo | 2.9 | 3.8 | 3.7 |

Table 3. Statistics of answers of participants in the five-point Likert scale.

| Evaluation | 1 | 2 | 3 | 4 | 5 |
|---|---|---|---|---|---|
| Operability of the user interface | 0 | 0 | 0 | 9 | 1 |
| Representation of the recommendation | 0 | 0 | 2 | 6 | 2 |

## 6.2 Evaluation by the photo set owner

We showed the tree structure and selected photos to the owner of the photo set, and recommended information setting the target region as "Okinawa" or "Hokkaido". Then, we asked the photo owner to subjectively evaluate the keyword selection result and representation of the tree structure. Also, we asked the satisfaction with the recommended sightseeing information.

The photo owner mentioned that the tree structure appropriately represented the categories of the targets of his travels. Meanwhile, he also mentioned that the granularity of keywords extracted by object recognition might be biased. For example, the object recognition process extracted a large number of names of animals. In other words, the selection of animal-related words was much finer than other categories of words. On the other hand, he mentioned that the set of words displayed by our user interface was well-balanced. Also, the photo owner pointed out that outdoor-related information is easy to retrieve because the word "outdoor" is selected as the root note, while indoor-related keywords are not sufficiently remarkable. This drawback might be due to our design that selects a single root node. We would like to extend the implementation so that we can select multiple root nodes and construct multiple trees.

Then, we asked the photo owner to observe the sightseeing information of "Okinawa" and "Hokkaido" retrieved by the presented system and answer how the information was satisfactory. He gave us the following comments with several particular words.

<u>beach, ocean:</u> Various beaches in Okinawa and Hokkaido were exhaustively introduced. It is convenient because we can easily select preferable beaches along with the planned drive courses.

<u>mountain:</u> Various mountains are exhaustively introduced. We can intuitively retrieve the sightseeing information by using the words related to the plan of the travels with the word "mountain."

The comments suggest several words extracted by Computer Vision API and displayed by our user interface are useful to retrieve sightseeing information. Meanwhile, we heard the following issues from the photo owner.

<u>plant:</u> Information related to botanical objects and manufacturing factories are mixed.

<u>nature:</u> retrieved information was too diverse because the word is abstract.

church: Information was not fruitful because the number of churches itself is small in Japan.

The above comments suggest that there are useful and less useful keywords for the retrieval of travel information. We would like to survey the satisfaction of the information retrieval results for each keyword so that we can preferentially show the particular keywords that may bring satisfactory retrieval results.

## 7 CONCLUSION AND FUTURE WORK

This paper presented a user interface to interactively retrieve sightseeing information based on the keywords extracted from the past photos of the users. The presented system applies generic object recognition APIs to the users' past travel photos and then constructs a tree structure of the keywords extracted from the photos. This paper focused on the user interface to show the tree structure and provide a mechanism to interactively select the query keywords and visualization of the retrieved sightseeing information. The paper also introduced the example results and user evaluation of the presented system.

We would like to improve the user interface based on the user evaluation results. Also, we would like to survey the satisfaction of the retrieval results for each keyword so that we can set the priority of the keywords.

Our graph clustering technique [6] has some limitations on this study. We need to automate the setting of the threshold to generate edges. Also, we would like to apply a dimension reduction technique to the input feature values because the feature values calculated from the frequency of all the keywords may be very high-dimensional and sparse.

After solving the above issues, we would like to conduct comparative experiments with existing techniques or commercial services on travel information retrieval or recommendation. Also, we would like to conduct experiments applying participants' own photos.

Currently, we suppose that users supply only their past travel photos because we would like to display keywords directly related to their past travels. In other words, many keywords not related to travels will be displayed if we allow users to supply all their personal photos including both travel and non-travel photos. Users may feel bothering if such unnecessary keywords are often displayed. Meanwhile, it is sometimes meaningful to include some kinds of non-travel photos as input photos. For example, it is possible that a user wants to visit a sports stadium during his/her next travel but he/she just has non-travel sports photos. In this case, non-travel sports photos may be useful for the user to explore keywords related to sports. Another issue is that it may be difficult to take photos at some kinds of travel spots such as department stores, public baths, and concert theaters. We cannot expect that we can extract keywords of such places from personal photos, and therefore we may need to additionally prepare such keywords. We would like to discuss the guideline on how to allow specific types of non-travel photos and what kinds of keywords we may need to add to the system.